\documentclass[prl,showpacs,twocolumn]{revtex4}

\usepackage[english]{babel}
\usepackage{graphicx}
\usepackage{bm}

\begin{document}

\title{Experimental teleportation of quantum entanglement with an optimal linear optics Bell-state analyzer}

\author{Thomas Jennewein$^1$}
\author{Rupert Ursin$^2$}
\author{Markus Aspelmeyer$^2$}
\author{Anton Zeilinger$^{1,2}$}
\affiliation{$^1$Institut f\"{u}r Quantenoptik und Quanteninformation, \"{O}sterreichische Akadmie der Wissenschaften, Boltzmanngasse 3, 1090 Wien, Austria}
\affiliation{$^2$Institut f\"{u}r Experimentalphysik, Universit\"{a}t Wien
Boltzmanngasse 5, 1090 Wien, Austria}

\date{\today}

\begin{abstract}
We demonstrate an experiment on entanglement swapping using an optimal Bell-state measurement capable of identifying two of the four Bell-states for polarization entangled photons, which is the optimum with linear optical elements. The two final photons belong to separately created paris. They are entangled after their original partner photons have been subjected to the Bell state measurement, whose outcome determines the type of entanglement of the final photon pair. The resulting violation of Bell's inequality in both cases confirms the success of the teleportation protocol.

\end{abstract}


\maketitle

Quantum state teleportation \cite{Bennett93a} allows the transfer of a quantum state from one system to an independent system, without obtaining any information on the state. This was first demonstrated with polarized photons in 1997 \cite{Bouwmeester97a}. A fascinating case of quantum teleportation is entanglement swapping \cite{Zukowski95a}, where the transferred state is not only unknown but undefined, since it is part of an entangled pair. According to quantum mechanics, the entanglement survives this procedure. In a recent  experiment\cite{jennewein02a}  we were able to show the quantum nature of the quantum teleportation, since the entanglement after the procedure exhibited a clear violation of Bell's inequality. The standard protocol for quantum state teleportation relies on two key resources which are the entanglement of particles and an entangling operation, called the Bell-state measurement. When using qubits in teleportation, there are four possible entangled states (Bell-states) and clearly it would be desirable to detect all four Bell-states, for the scheme to be perfectly efficient. Provably this is not possible with linear optical elements \cite{Calsamiglia01a} and only two of the four Bell-states can be identified perfectly. This Bell-state analysis was demonstrated experimentally in quantum dense coding\cite{Mattle96a,Michler96a}.  Yet, in  previous experiments on entanglement swapping with photons\cite{jennewein02a}, a simple Bell-state analyzer was implemented which was only able to detect one of the four Bell-states. Here we describe the first realization of teleportation of entanglement where an enhanced Bell-state analyzer (BSA) was used. The newly implemented BSA was capable of detecting two of the four Bell-states, and its actual result for each successful teleportation event determined the correlation of the teleported photon. This setup allowed us to observe that the swapped entanglement violated Bell's inequality separately for each of the two possible Bell-states obtained in the BSA. In a parallel experiment \cite{rupert} the optimal linear Bell-state analyzer was employed in a conventional quantum teleportation experiment, where the polarization state of individual photons was teleported.

The setup of our system is shown in Fig.~1, which is based on the system shown in \cite{jennewein02a}. The Bell-state analyzer was realized by interfering the two photons $b$ and $c$ on a 50:50 beam splitter (BS), and with additional polarizers (PBS) in each of the output arms of the BS. The detection of the two  Bell-state went as follows: when one photon is detected in each output arm of the beam splitter BS, and in opposite outputs of
their polarizers, then a $|\Psi^-\rangle_{bc}$-Bell-state was observed (i.e. either detectors D1H
and D2V fire, or D1V and D2H). This is the conventional Bell-state analysis with photons. However, if the two photons were detected in the same output
arm of the beam splitter , but in different outputs of the polarizer, then a $|\Psi^+\rangle_{bc}$-Bell-state was observed (either D1H and D1V fire, or D2H and D2V).  The interference performed within this BSA required full compensation of the polarization rotation in all the fibers, which was a rather tedious and timely process. However, once good alignment was achieved, the polarization in the fibers was stable to about 5$^\circ$ over 24~h. The rate of teleportation events achieved in this setup was about 0.0065 per second. A measurement for each setting of the polarizers lasted over $10000$ seconds.

The emission characteristics of our photon pair production implies a similar
probability for producing two photon pairs in separate modes (one photon each
in modes $a, b, c, d$) or two pairs in the same mode (two photons each in modes
$a$ and $b$ or in modes $c$ and $d$). The latter events could also lead to ``false" coincidences in
the detectors in the BSA. We exclude these cases by only
accepting events where one photon each in mode~$a$ and mode~$d$ is registered.
However, our optimal Bell-state analyzer diminishes this problem, since it reduces the
unwanted two-pair events by at least a factor of two. Only those cases are detected in the
Bell-state analyzer, when the two photons ($b$ and $c$) take separate $H,V$ outputs of the
polarizing beam splitters arranged behind the beam-splitter.

To show the presence of entanglement after the teleportation, we test the correlations between photons $a$ and  $d$ with a Bell-inequality, where a suitable version for experiments is the  Clauser-Horne-Shimony-Holt (CHSH) variant \cite{Clauser69a}. This inequality uses a fair sampling hypothesis to overcome
the inherent losses in an experimental system. Its evaluation requires a total
of four correlation measurements performed on pairs of particles with
different analyzer settings. The CHSH inequality has the following form:
\begin{equation}
	S=|E(\phi_a',\phi_d')-E(\phi_a',\phi_d'')|+|E(\phi_a'',\phi_d')+E(\phi_a'',\phi_d'')|\leq2, \label{chsh}
\end{equation}
where $S$ is called the ``Bell parameter'' and $E(\phi_a,\phi_d)$ are the correlation
coefficients for polarization measurements where $\phi_a$ and $\phi_a$ are the polarizer
setting for photon~$a$ and  photon~$d$ respectively\cite{Aspect82b}.The correlation coefficients are defined as
\begin{eqnarray}
	E(\phi_a,\phi_d)&=&(N_{++}(\phi_a,\phi_d)-N_{+-}(\phi_a,\phi_d)\\ \nonumber
	& & -N_{-+}(\phi_a,\phi_d)+N_{--}(\phi_a,\phi_d))/ \sum N_{ij}
\end{eqnarray}
where as usual the
$N_{ij}(\phi_a,\phi_d)$ are the coincidence counts between the $i$--channel of
the polarizer of photon~$a$ set at angle $\phi_a$, and  the $j$--channel of the
polarizer of photon~$d$ set at angle $\phi_d$.It is straight forward to show that the prediction of quantum mechanics leads to $E^{QM}(\phi_a,\phi_d)=-\cos(2(\phi_a-\phi_d))$. With a choice of settings such as
$(\phi_a',\phi_d',\phi_a'',\phi_d'') = (0^\circ,
22.5^\circ,45^\circ,67.5^\circ)$ then leads to the maximal value of
$S^{QM}=2\sqrt{2}$, which clearly violates the limit of $2$ imposed by inequality~(\ref{chsh}) and hence leads to a
contradiction between local realistic theories and quantum mechanics
\cite{Bell64a}.

A measurement of the correlation coefficients $E(\phi_a,\phi_d)$ for photons~$a$ and $d$ at settings
$\phi_a=\phi_d=45^\circ$ is shown in Figure~2, where the time delay between the
two photons~$b$ and $c$ interfering in the Bell-state analyzer was varied.
Clearly the dependance of the photon interference on the delay between the two photons can be observed and highlights the operation of the interferometric Bell-state analysis used in this entanglement
swapping experiment. Also this measurement shows the very high correlation of the entagled photons after the teleportation of $0.91$ and $0.86$ for the $|\Psi^-\rangle_{ad}$ and $|\Psi^+\rangle_{ad}$ respectively. Note that this correlation includes all the degrading effects of the teleportation procedure itself as well as the initial quality of the entangled state input. A detailed study of the different effects which reduce the teleportation quality is given in \cite{jennewein02b}. 

For a test of Bell's inequality, a series of correlation coefficients of
photon~$a$ and photon~$d$ were measured, and are given in table~1. Note that the correlation coefficients for the two Bell-states were measured within the same
measurement run for a particular polarizer setting, however by sorting the
data into corresponding subsets depending the observed Bell-state. The results show clear violations of the Bell-inquality since the Bell-parameter is $S=2.60 \pm 0.09$ for the $|\Psi^-\rangle_{ad}$ and $S=2.30 \pm 0.09$ for  the $|\Psi^+\rangle_{ad}$ Bell-state. The lower value of $S$ for $|\Psi^+\rangle_{ad}$ is due to the difficult polarization alignment of the output fibers of the beam splitter for this case.

We demonstrated for the first time the teleportation of entangled photons using an optimal Bell-state analyzer capable of detecting two of the four Bell-states perfectly. Our system achieved high visibilities and therfore enabled the teleported entangled states to violate two Bell-inequalities in the same experiment, by sorting the polarization results of the teleported photons into subsets corresponding to the two-fold output of the  Bell-state analyzer. This violations of the Bell-inequalities in our optimal entanglement swapping experiment additionally highlights the non-classical nature of entanglement as the outcome of the Bell-measurement of the two inner photons~$b$ and $c$ ($|\Psi^-\rangle_{bc}$ and $|\Psi^+\rangle_{bc}$) even determines the type of entanglement of the two outer photons~$a$ and $d$ ($|\Psi^-\rangle_{ad}$ and $|\Psi^+\rangle_{ad}$). 

It is surprising to note, that our teleportation scheme based on the two-fold Bell-state analysis is also optimal in a more general view. One would naively expect that the teleportation of entanglement is perfect for all four Bell-states, even when these are non-maximal entangled, as would generally be the case in a practical system.  Yet, Bose, Vedral and Knight\cite{Bose99a} were able to show that two of the four Bell-states will experience a purification of the final entanglement, whereas the other two Bell-states will become even less entangled. A related experiment was carried out recently\cite{jian_wei}. The consequence of this insight is that entanglement swapping realized with a full Bell-state analyzer would not represent a significant advancement over the experiment described here, for systems containing non-perfectly entangled states.
 
Our experiment demonstrated the features of quantum state teleportation in a very fascinating context since the actual correlation between two distant particles depends on the outcome of the Bell-state measurement on their entangled partners. Moreover, since teleportation using a two-fold Bell-state measurement is also one of the building blocks of linear-optical quantum gates\cite{Knill00a}, this experiment  also represent a step towards all-optical quantum computation.

This work has been supported by the Austrian Science Foundation
(FWF) Project No. F1506 and F1520, and by the European Commission, Contract
No. IST-2001-38864 (RamboQ). M.A. has been supported by the
Alexander von Humboldt foundation. 




\begin{table}
\centering
\begin{tabular}{c|c|c} \hline
  $E(\theta_a;\theta_d)$ & $|\Psi^-\rangle_{ad}|\Psi^-\rangle_{bc}$
  & $|\Psi^+\rangle_{ad}|\Psi^+\rangle_{bc}$ \\ \hline
  $E(0^\circ; 67.5^\circ) $ & $-0.80 \pm 0.04$ & $-0,60 \pm 0,04$ \\
  $E(0^\circ; 22.5^\circ) $ & $-0.61 \pm 0.04$ & $0.55 \pm 0.04$ \\
  $E(45^\circ; 67.5^\circ) $ & $0.64 \pm 0.05$ & $-0.48 \pm 0.05$ \\
  $E(45^\circ; 22.5^\circ) $ & $0.55 \pm 0.05$ & $-0.67 \pm 0.05$ \\
  \hline \hline
  Bell-Parameter $S$ & $2.60 \pm 0.09$ & $2.30 \pm 0.09$ \\
  \hline
\end{tabular}
\caption{Violation of the CHSH-inequality with the measured correlation coefficients between photons~$a$ and $d$ for the
suitable polarizer settings $\theta_a;\theta_d$.  The outcome of the Bell-state
measurement on photons~$b$ and $c$ determines the type of correlation. As the local realistic upper limit for $S$ is 2, above results
for both Bell-states clearly violate the CHSH-inequality. Note that the
correlation coefficients for the two Bell-states were measured within the same
measurement run for a particular polarizer setting, however by sorting the
data into corresponding subsets. The given errors are the statistical
uncertainties due to photon counting. The differences
in the correlation coefficients come from the higher correlation fidelity for
analyzer settings closer to $0^\circ$ and $90^\circ$.}
\end{table}

\begin{figure}[h]

\includegraphics[width=13cm]{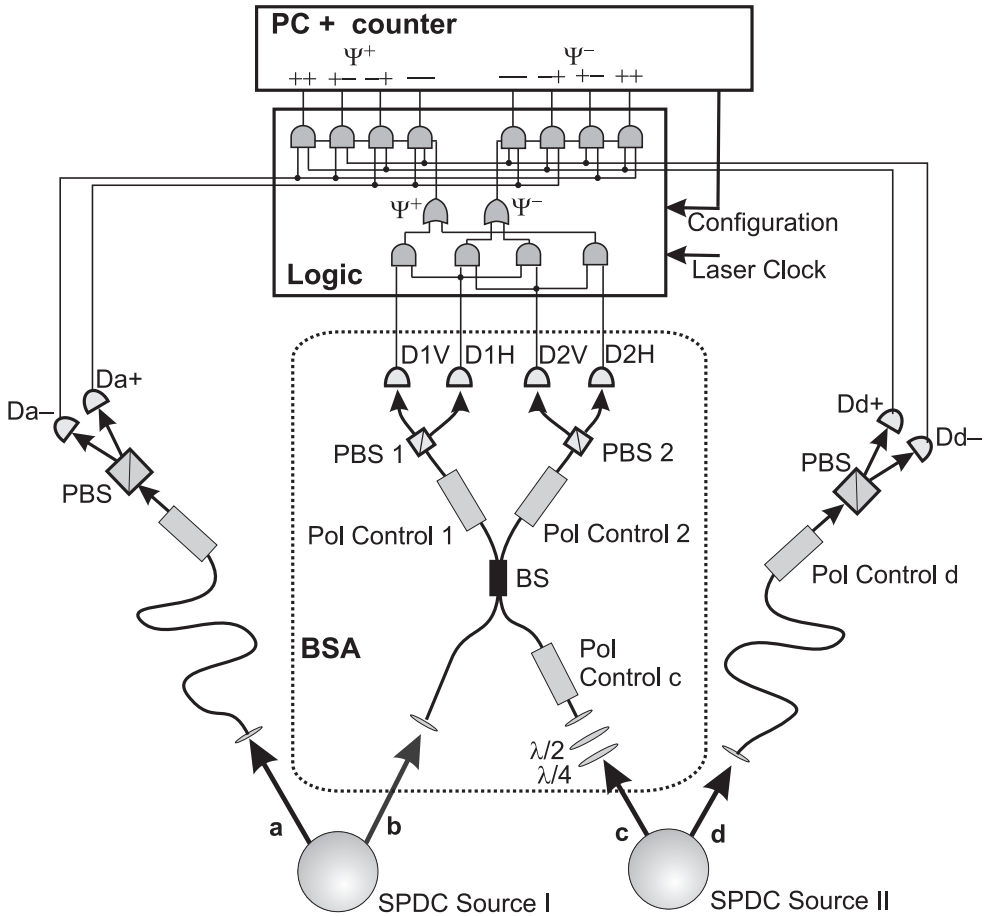}
\caption{Setup of entanglement-swapping experiment using optimal Bell-state analysis. The two photon pairs are produced by spontaneous parametric down conversion\cite{Kwiat95b} in barium-borate (BBO) pumped by UV-laser pulses (wavelength=394~nm, width=200~fs, rate=76~MHz).   From the two entangled photon pairs
(pair~$a-b$, and pair~$c-d$), one photon each (photons~$b$ and $c$) is
sent to  the Bell-state analyzer (BSA), consisting of a 50:50 fiber beam splitter (BS)
and with polarizing beam splitters (PBS) in each output arm. The temporal overlap between photons~$b$ and $c$ is adjusted by moving the mirror for the UV-pulses.  The remaining photons~$a$ and
$d$ are separately analyzed with polarizers (PBS) and detected in each output of the PBS. The spurious birefringence of the fibers which induces polarization rotations of the photons is compensated with polarization controllers. For the fiber beam splitter, also an additional halfwaveplate ($\lambda/2$) and quarterwaveplate ($\lambda/4$) are needed for the compensation of the phase between the reference polarization states. A coincidence logic compares all eight detector signals and identifies the desired useful events of the four photons, as sketched in the figure, with boolean AND and OR gates. The logic was realized in a programmable logic array \cite{jennewein02b}. The occurrences of the eight possible four-fold events were counted in a PC-counter-card. The logic also included an adjustment mode for counting the two-folds and the singles count rates (configured from the PC). }

\end{figure}

\begin{figure}[h]

\includegraphics[width=8cm]{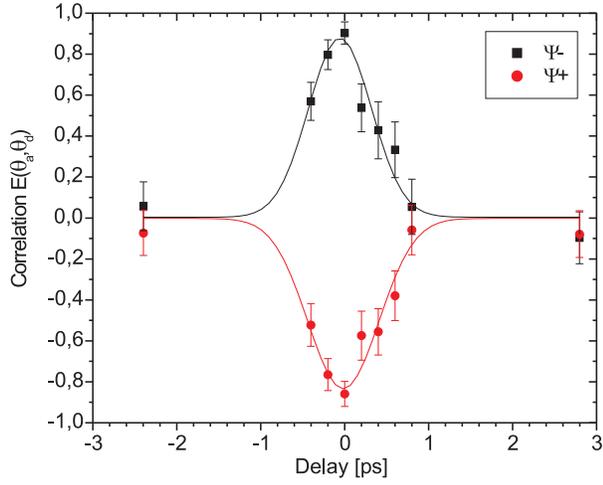}
\caption{Measurement of the correlation coefficient for photons~$a$ and $d$,
depending on the relative delay of photons~$b$ and $c$ interfering in the
Bell-state analyzer. The setting of the polarizers was $\theta_a=\theta_d=45^\circ$.  The maximal correlation occurs for the optimal temporal overlap
of the two wave packets of photons~$b$ and $c$. The two correlations were obtained in the same measurement
by sorting the detection events in respect with the outcome of the
Bell-state measurement. }
\end{figure}

\end{document}